\documentclass[aps,prd,twocolumn,nofootinbib,floatfix]{revtex4-1}
\usepackage{setspace}
\usepackage{amsthm}
\theoremstyle{definition}

\usepackage{graphicx}
\usepackage{dcolumn}
\usepackage{multirow}
\usepackage{subfigure}
\usepackage{times,mathptm}
\usepackage{float}
\usepackage{color}
\usepackage{amsmath,amsfonts}
\usepackage{mathptmx}
\usepackage{mathrsfs}
\usepackage{bbm}
\usepackage{bm}
\usepackage{xfrac}
\newcommand{\beq}{\begin{equation}}
\newcommand{\eeq}{\end{equation}} 
\newcommand{\bea}{\begin{eqnarray}}
\newcommand{\eea}{\end{eqnarray}} 

\newcommand{\A}{\mathcal{A}}
\newcommand{\x}{\mathbf{x}}

\newcommand{\T}{\mathcal{T}}

\newcommand{\E}{\mathcal{E}}

\renewcommand{\l}{\lambda}

\renewcommand{\b}{\beta}
\renewcommand{\a}{\alpha}

\newcommand{\vx}{{\vec{x}}}
\newcommand{\vy}{{\vec{y}}}
\newcommand{\vz}{\vec{z}}

\newcommand{\pbar}{\overline{\psi}}

\newcommand{\g}{\gamma}

\newcommand{\D}{\Delta}

\renewcommand{\th}{\theta}

\newcommand{\oh}{\frac{1}{2}}

\newcommand{\dg}{\dagger}
\newcommand{\non}{\nonumber}
\renewcommand{\t}{\tau}
\newcommand{\rf}[1]{(\ref{#1})}
\newcommand{\ra}{\rightarrow}
\newcommand{\pa}{\partial}
\renewcommand{\vec}[1]{\bm #1}
\usepackage{ulem}

\begin{document}

\title{Quantum excitations of static charges in the Ginzburg-Landau model of superconductivity} 

\bigskip
\bigskip

\author{Jeff Greensite and Kazue Matsuyama}
%\singlespacing
\affiliation{Physics and Astronomy Department \\ San Francisco State
University  \\ San Francisco, CA~94132, USA}
\bigskip
\date{\today}
\vspace{60pt}
\begin{abstract}

\singlespacing
 
     We point out that  in superconductors there may exist localized quantum excitations of the electric and condensate fields surrounding a static charge, which cannot be interpreted as simply the ground state of the screened charge plus some number of massive photons.  This is illustrated via a lattice Monte Carlo calculation of the energy spectrum of a pair of separated static charges in an effective Ginzburg-Landau model of superconductivity.

\end{abstract}

\maketitle
 
\singlespacing

\section{\label{Intro} Introduction}

   Composite quantum systems typically have a discrete spectrum of excitations, and although certain objects  such as electrons and
muons are referred to as ``elementary'' particles, this terminology can be a little misleading, since in any interacting quantum field theory these elementary particles are inevitably surrounded by a field of some kind.  In this sense, elementary particles are also composite. The question we would like to address here is whether there can exist a discrete spectrum of localized excitations in the fields surrounding a static charge, and in particular whether this might be the case for the fields surrounding static ions in a superconductor.

    Let us begin with the simplest case: a static charge coupled to the free quantized Maxwell field.  Gauge invariance of physical states under infinitesimal gauge transformations, embodied in the Gauss Law, insists that the static charge must be the source of an electric field.   The ground state of the static charge + gauge field system, as shown long ago by Dirac \cite{Dirac:1955uv}, is given by
\bea
     |\Psi_\vx \rangle = \pbar(\vx) \rho(\vx;A) |\Psi_0 \rangle \  ,
\label{qed}
\eea
where $\pbar(\vx)$, operating on the vacuum, creates the static charge, and
\bea
     \rho(\vx;A) &=& \exp\left[-i {e\over 4\pi} \int d^3z ~ A_i(\vz) {\pa \over \pa z_i}  {1\over |\vx-\vz|}  \right] \  .
 \label{rho}
\eea
In the state $|\Psi_\vx \rangle $ the expectation value of the electric field is simply the Coulomb field of the static charge.  The operator $\rho(x;A)$, which depends only on the gauge field, is an example of what we call a ``pseudomatter'' operator.{\footnote{It is worth noting that $\rho^*(x;A)$ is also the gauge transformation to Coulomb gauge.  In fact, if
 $g(x;A)$ is the gauge transformation to a physical gauge defined by some condition $F[A]=0$ imposed on each time slice, then it is not hard to see that $g^*(x;A)$ is pseudomatter operator. This statement can be readily generalized to non-abelian theories.}  Operators of this kind, functionals of only the
gauge field, transform under an infinitesimal gauge transformation like a static charge, but are invariant under global gauge transformations $g(x)=g$ in the center of the gauge group, which in this case is $g(x) = \exp(i\th) \in$ global U(1). As a result, the physical state \rf{qed} is not {\it entirely} gauge invariant; in fact it transforms covariantly under such global U(1) transformations, i.e.
\bea
|\Psi_\vx \rangle \ra e^{-i\theta} |\Psi_\vx \rangle \ . 
\eea
This is characteristic of an isolated charge in electrodynamics, unscreened by any matter fields.  It should be
emphasized that covariance under the global U(1) subgroup of the gauge group is in
no way a violation of the physical state condition (i.e.\ the Gauss law), which does not require invariance
under global transformations which affect neither the gauge field nor the charge densities.   Of course this global symmetry of the action may or may not be a symmetry of the vacuum. As we have discussed extensively elsewhere \cite{Greensite:2020nhg,*Greensite:2021fyi}, it is the global center subgroup of the gauge group which is broken spontaneously in the Higgs phase of a gauge Higgs theory, with a corresponding gauge-invariant order parameter closely analogous to the Edwards-Anderson order parameter \cite{Edward_Anderson} for a spin glass.  Symmetry breaking of this kind is not a violation of Elitzur's theorem, which forbids the spontaneous breaking of local, rather than global symmetries. While the transition to the Higgs phase is not necessarily a thermodynamic transition, it is nonetheless accompanied by a change in physical behavior, which is either the loss of metastable flux tube states and Regge trajectories (transition from the confinement phase), or the disappearance of massless vector bosons (transition from the Coulomb phase).

   For our purposes the free field example is not so interesting.  Of course there are excitations of the charge+quantized field combination, but these excitations consist simply  of some number of photons, of arbitrary positive energies, superimposed on a background Coulombic field; they are not the localized excitations with a discrete energy spectrum that are of interest here.  We therefore move on to gauge Higgs theories, including those which, like the Ginzburg-Landau model, may simply be the effective theory of some more fundamental physics  in which the Higgs field is composite.  There is already evidence of excitations of static charges in SU(3) gauge Higgs theory \cite{Greensite:2020lmh}, and in the abelian Higgs model 
\cite{Matsuyama:2020tvt}.  But the theories of phenomenological interest would be the Ginzburg-Landau model of superconductivity, and the electroweak sector of the standard model.  A spectrum of quark and lepton excitations, or more precisely a spectrum of localized excitations of the fields surrounding quarks and leptons, would obviously be of great interest, but here we face the formidable complication that the electroweak theory is a chiral gauge theory, which resists a lattice formulation.\footnote{See, however, the attempt to find excitations in a chiral U(1) theory in \cite{Greensite:2021jef}.}  For that reason we turn our attention to quantum behavior in the Ginzburg-Landau model with static ions.
   
\section{the calculation}

The effective Ginzburg-Landau action is (cf.\ \cite{coleman_2015})%,*altland_simons_2010}
\bea
S &=& \int d^4x \bigg\{ \oh \rho_s \bigg({1 \over \upsilon^2} (\pa_0 \xi + 2eA_0)^2  \non \\
& & \qquad  - (\nabla \xi -2e{\bf A})^2 \bigg)  + \oh(E^2 - B^2) \bigg\} \ ,
\eea
where $\xi$ is the Goldstone mode, $\rho_s=n_s/2M$ where $n_s, M, 2e$ are the Cooper pair density, mass, and charge respectively, and $\upsilon \sim 10^{-2}$ is on the order of the ratio of the Fermi velocity in a metal to the speed of light.   The mass of the transverse photon in the continuum formulation is, in natural units,
\beq
           M_{ph} = {1\over \lambda_L} = 2e \sqrt{\rho_s} \ ,
\label{tree}
\eeq
where $\lambda_L$ is the London penetration depth.
The transition to the lattice formulation in Euclidean time, which is the non-relativistic version of the charge $q=2$ lattice abelian Higgs model, is straightforward:
\bea
S_{eff} &=& -\beta \sum_{plaq} \text{Re}[UUU^*U^*] \non \\
& & - \gamma  \sum_x \sum_{k=1}^3 \text{Re}[\phi^*(x)U^2_k(x)\phi(x+\hat{k}) ]  \non \\
& &  - {\gamma \over \upsilon^2}  \sum_x \text{Re}[\phi^*(x)U^2_0(x)\phi(x+\hat{t}) \ ,
\eea 
where $\beta=1/e^2 \approx10.9, ~ \phi=e^{i\xi}, ~ \rho_s=\gamma/a^2,$ and $a$ is the lattice spacing.  A choice of gamma, together with an observed London penetration depth and the tree-level relationship \rf{tree}, requires a lattice spacing
\beq
      a = 2e \gamma^\oh \lambda_L \ .
\label{spacing}
\eeq
Obviously the continuum limit would require $\g \ra 0$, but this is impossible because of a transition to the
massless phase at finite $\g$.  At $\b=10.9$ we find, from investigation of the susceptibility of 
$\langle \text{Re}[\phi^*(x)U^2_k(x)\phi(x+\hat{k})\rangle$, a transition close to $\g=0.017$ (see the Appendix).  So it must be emphasized that the lattice formulation is an effective theory, valid beyond some inherent short distance cutoff which, since Cooper pairs are composite, would most naturally be the diameter of a Cooper pair.   In most texts this effective theory is treated classically, but here we would like to treat the effective theory as a quantum theory in its own right, with a short distance cutoff $a$.  In lattice Monte Carlo simulations of $S_{eff}$ it is convenient to fix to unitary gauge, in which case , since $1/\upsilon^2 \sim 10^4$, it is a very good approximation to take $U_0(x) = \pm 1$.  

The aim is to search for excitations
around pairs of widely separated static $q=\pm 1$ (e) charges, having fixed ions in mind.  We are obliged
to consider pairs of charges for technical reasons discussed below, which are related to the fact that
one cannot create a single charge in a periodic volume. We therefore consider physical states of the following
form:
\beq
|\Phi_n(\vx,\vy) \rangle = Q_n(\vx,\vy) |\Psi_0\rangle \ ,
\label{Pn}
\eeq
with
\bea
    Q_{2n-1}(\vx,\vy) &=& 
               \pbar(\vx) \zeta_n(\vx) \zeta_n^*(\vy) \psi(\vy) \non \\ 
   Q_{2n}(\vx,\vy) &=&             
               \pbar(\vx) \phi(\vx) \zeta^*_n(\vx) \zeta_n(\vy) \phi^*(\vy) \psi(\vy)  \ ,
\label{Qn}
\eea
where $\pbar,\psi$ are static fermion operators, and the pseudomatter operators $\zeta_n(x)$ are eigenstates of the covariant Laplacian operator 
\beq
       \sum_{\vy}  (-D^2)_{\vx \vy} \zeta_n(\vy) = \lambda_n \zeta(\vx) \ ,
\eeq
where
\bea
  (-D^2)_{\vx \vy} =  \sum_{k=1}^3 \left[2 \delta_{\vx \vy} - U_k(\vx) \delta_{\vy,\vx+\hat{k}} 
       - U_k^{\dg}(\vx-\hat{k}) \delta_{\vy,\vx-\hat{k}}  \right] \ . \non \\
\label{D2}
\eea 
Note that $\zeta_n(\vx)$ is a pseudomatter operator, and $\phi(\vx)\zeta_n^*(\vx)$ transforms like a pseudomatter operator.  We have found it important,
in the calculations discussed below, to include both.

    Now let us consider a subspace of Hilbert space spanned by $N$ physical states 
$\{|\Phi_n(\vx,\vy)\rangle, n=1,..,N\}$, and let $\tau$ be the transfer matrix.  Define the rescaled transfer matrix
\beq
\T = \t e^{\E_0} \ ,
\eeq
where $\E_0$ is the vacuum energy (or, more precisely, if $\kappa_0$ is the largest eigenvalue of $\t$,
then $\E_0 = -\log(\kappa_0)$). We would like to construct a set of eigenstates of $\T$ in the
truncated basis.  
For this non-orthogonal basis, we first calculate numerically the matrix elements and overlaps,
\bea
          [\T]_{\alpha \beta}(R) &=& \langle \Phi_\alpha | \T | \Phi_\beta \rangle  \non \\
                                            &=& \langle Q^\dg_\a(\vx,\vy,1) Q_\b(\vx,\vy,0) \rangle \non \\
          \left[O\right]_{\alpha \beta}(R) &=& \langle \Phi_\alpha | \Phi_\beta \rangle \non \\
                                             &=& \langle Q^\dg_\a(\vx,\vy,0) Q_\b(\vx,\vy,0) \rangle  \ .   
\eea  
where $R=|\vx-\vy|$, and $Q(\vx,\vy,t)$, in Euclidean time path-integral formulation, is the $Q(\vx,\vy)$ operator acting at time $t$.  We obtain the eigenvalues of $\T$ in the subspace by solving the generalized eigenvalue problem,
\bea
         [\T] \vec{\upsilon}^{(n)} &= &\lambda_n [O] \vec{\upsilon}^{(n)} \non \\
          |\Psi_n(\vx,\vy)\rangle &=& \sum_{\a=1}^N \upsilon^{(n)}_\a |\Phi_\a(\vx,\vy)\rangle \ .
\label{Geig}
\eea
Each $|\Psi_n(\vx,\vy)\rangle$ is a linear combination of the non-orthogonal basis states $|\Phi_\a(\vx,\vy)\rangle$, and the set of states $\{|\Psi_n(\vx,\vy)\rangle\}$ are the energy eigenstates (i.e.\ eigenstates of the transfer matrix) of the isolated static pair in the subspace.  

   Let us define, for $M$ pseudomatter operators $\zeta_\a$, the $N=2M$ operators
\bea
      V_{2\a-1}(\vx,\vy,U) &=& \zeta_\a(\vx) \zeta^*_\a(\vy) \non \\
      V_{2\a}(\vx,\vy,U) &=&   \zeta^*_\a(\vx) \phi(\vx) \phi^*(\vy) \zeta_\a(\vy)  
\eea
In order to calculate the required matrix elements numerically, we integrate
out the heavy fermions (via a hopping parameter expansion \cite{Gattringer:2010zz}) to obtain
\bea
 [\T^T]_{\alpha \beta}(R)  &=& \langle \Phi_\a|\T^T |\Phi_\b\rangle \non \\
  &=& \langle Q_\a^\dg(\vx,\vy,T) Q_\b(\vx,\vy,0) \rangle \non \\
   &=&  \langle V^\dagger_\alpha(\vx,\vy;U(t+T)) P^\dagger(\vx,t,T)  \non \\ 
    & & ~~ \times V_\beta(\vx,\vy;U(t)) P(\vy,t,T)  \rangle \non \\
     \left[O\right]_{\alpha \beta}(R) &=&    \langle V^\dagger_\alpha(\vx,\vy;U(t))  V_\beta(\vx,\vy;U(t)) \rangle
\label{transf2}
\eea
where indices $\a,\b$ range from 1 to $N$, and
\beq
          P(\vx,t,T) = U_0(\vx,t) U_0(\vx,t+1)...U_0(\vx,t+T-1) \ .
\eeq
is a timelike Wilson line of length $T$.  In \rf{transf2} we have dropped powers of the fermion mass in the hopping parameter expansion, which only contribute an overall constant to the energy of the fermion-antifermion system, and are therefore irrelevant to the question of excitations.

   The eigenstates $\{|\Psi_n\rangle\}$ of the transfer matrix in the $N$-dimensional subspace are in general not eigenstates of the transfer matrix in the full Hilbert space.  But it may happen that $\Psi_1$, the state
with the lowest energy expectation value in the subspace, has a very large overlap with the static charge ground state
in the full Hilbert space, and it follows that the $\Psi_{n>1}$ would have a correspondingly small
overlap.  On general grounds
\bea
          \T_{nn}(R,T) &\equiv& \langle \Psi_n|\T^T|\Psi_n \rangle \non \\
         &=& \sum_{\a \b} \upsilon^{*(n)}_\alpha  \langle Q_\alpha^\dagger(\vx,\vy,T) Q_\beta(\vx,\vy,0) \rangle \upsilon^{(n)}_\beta \non \\
                      &=& \sum_k |c_k(R)|^2 e^{-E_k(R) T} \ ,
\eea
where $E_k$ is the energy above the ground state of the ${k\mbox{-th}}$ energy eigenstate in the full Hilbert space,
containing a static fermion-antifermion pair separated by a distance $R$.  If $\Psi_{n>1}(\vx,\vy)$ has a
large overlap with one excited energy eigenstate $\Psi^{exact}_i$, and a very small overlap with the ground state, then we may expect that for some range of $T_{min} \le T\le T_{max}$
\beq
           \T_{nn}(R,T) \approx |c_i(R)|^2 e^{-E_i(R) T} ~~~,~~~ T_{min} \le T\le T_{max} \ ,
\eeq
and in that case we may extract the excitation energy $E_i(R)$ from a logarithmic plot of $\T_{nn}(R,T)$ vs.\ $T$.

   Of course there is no guarantee that a strategy of this kind will work.  It relies
on the conjectures that the overlap of $|\Psi_1\rangle$ with the true $\pbar \psi$ ground state is very large. 
Even if this conjecture is true, it might be the case that the excited states are nothing more than the
ground state plus one or more gauge bosons, as in pure QED.  But the general idea is testable, and our results for the Ginzburg-Landau model in the Higgs phase are reported in the next section.

\subsection{Why not single fermion states?}

   Before presenting numerical results, we would like to address this question: why not consider single
fermion states of the form
\beq
           \Omega_\a = \pbar(\vx) \zeta_\a(\vx) \Psi_0  \ ,
\eeq
where $\zeta_\a$ is a pseudomatter field.   We assume that there is some algorithm which determines 
$\zeta(\vx,t)$ uniquely on time slice $t$ from the spacelike link variables on that time slice.  Such numerical algorithms exist for Laplacian eigenstates, and also for numerical fixing to Coulomb gauge.\footnote{The iterative gauge-fixing algorithms used in computer simulations are deterministic, and fix to a unique Gribov copy satisfying the gauge fixing condition.}  The latter determines the gauge transformation to Coulomb gauge, which is itself a pseudomatter field, in a finite periodic volume. 

    Now consider the transfer matrix elements 
\beq
       \langle \Omega_\a|\T|\Omega_\b\rangle = \langle \zeta_\a^\dg(\vx,1) U^\dg_0(\vx,0) \zeta_\b(\vx,0) \rangle  \ ,
\label{vev}
\eeq
which is essential in our discussion of energy expectation values.  The problem here is that the operator on the right hand side of \rf{vev} is not invariant under gauge transformations $g(\vx,t) = e^{i\th(t)}$ or, in an SU(N) gauge theory, under transformations
$g(\vx,t) = z(t) \in Z_N$.  Transformations of this kind do not alter spacelike links, and hence do not alter the pseudomatter fields
$\zeta_\a(x)$.  Nor do these transformations, which in each time slice belong to the global $Z_N$ center subgroup of the gauge group, alter the gauge-invariant action. But they
{\it do} transform the timelike links $U_0(\vx,t)$.  As a result, the matrix elements \rf{vev} all vanish,
at least in a finite periodic volume.  This is very general, it holds both in pure gauge theory and gauge Higgs theories, and it is one way of understanding why, on the lattice, it is impossible to place a single charge in a periodic volume, despite the fact that one can numerically transform the gauge field to
some unique Gribov copy of Coulomb gauge.  If there were a single-charged matter field in the theory, 
call it $\varphi$, then the same  objection would not hold for the neutral state
\beq
           \Psi = \pbar(\vx) \varphi(\vx) \Psi_0  \ ,
\eeq
 as the operator $\pbar(\vx) \varphi(\vx) $ is fully gauge invariant, including under global gauge transformations, and
 the expression in \rf{vev}, with $\zeta$ replaced by $\varphi$, is invariant under $g(\vx,t) = e^{i\th(t)}$.
 But there are no such single-charged dynamical fields in the Ginzburg-Landau model under consideration.
 
    It is for these reasons that we must consider pseudomatter states with two static charges, \
 \bea
      \Phi_{\a \b}(\vx,\vy) = \pbar(\vx) \zeta_\a(\vx) \zeta^*_\b(\vy) \psi(\vy) \Psi_0  \ ,
 \eea
 where the $\zeta_\a$ are eigenstates of the covariant Laplacian operator.  But here there is still
 an ambiguity, in that if $\zeta_\a(\vx)$ is a Laplacian eigenstate, then so is
 \beq
           \zeta'_\a(\vx;U) = e^{iF_\a[U]} \zeta_\a(\vx;U)  \ ,
  \label{ambig}
  \eeq
  where $F_\a[U]$ is any real-valued functional of the link variables.  So in general we may expect that any
  $\zeta_\a(\vx;U)$ determined by the Arnoldi algorithm will have wild fluctuations in a global phase factor
  as $U$ is varied, and fluctuations of that sort will cause
  \beq
             \langle \Phi_{\a\b}|\T^T|\Phi_{\a\b}\rangle
 \eeq
 to vanish, for any $T>0$, and $\a\ne \b$.  And in fact this is what is observed in numerical simulations.  For this reason
 we must restrict our considerations to states $\Phi_n$ of the form  (\ref{Pn},\ref{Qn}), where there is no
 remaining ambiguity, and  any global phase factor as in \rf{ambig} cancels out.
 
 \subsection{Photon mass}
 
 The photon mass in the Ginzburg-Landau model is determined from the time correlation of space-averaged gauge-invariant link operators.
 We define
 \beq
          G(t) = {1\over 3}\sum_{i=1}^3 \langle \A_i(0) \A_i(t) \rangle  \ ,
 \eeq
 where
 \beq
          \A_i(t) = {1\over L^3} \sum_\x \mbox{Im}[\phi^\dg(\vx,t) U^2_i(\vx,t) \phi(\vx + \hat{i})]   \ .
 \eeq
 A typical result, at $\b=10.9$ and $\g=0.60$ is shown in  ${\mbox{Fig.~} \ref{photon}}$.  The data was taken on a $16^3 \times 36$ lattice volume, with 400 lattices separated by 100 sweeps.  An exponential fit in the range $1 \le t \le 10$ gives a
 value of $m_{ph} = 0.446(3)$ for the photon mass in lattice units.  A similar fit with the same parameters, except for
 $\g=0.25$, results in $m_{ph} = 0.288(1)$.  We note that both of these values are quite close to the tree-level
 expression $m_{ph} = 2e\sqrt{\g}$ (with $m_{ph}$ in lattice units), obtained by multiplying both sides of eq.\ \rf{tree}
 (where $M_{ph}$ is in physical units) by the lattice spacing $a$.
          
 \begin{figure}[htb]
 \includegraphics[scale=0.6]{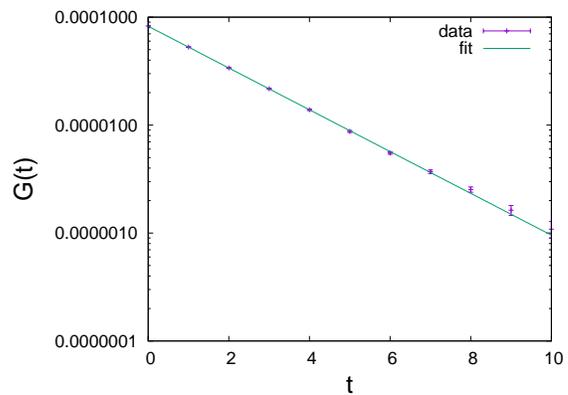}
 \caption{Determination of the photon mass at $\b=10.9, \g=0.6$ from the time correlator $G(t)$ of space averaged
 gauge invariant link variables on a $16^3 \times 36$ lattice volume.}
 \label{photon}
 \end{figure}
          
 \section{Excitations}
 
    We set $\b=10.9$, corresponding to ${e^2/4\pi} =1/137$ in natural units, and begin with Higgs coupling $\g=0.6$ on
a $12^3 \times 36$ lattice volume.  We consider static charge $q=\pm 1$ pairs at all pairs of lattice coordinates  $\vx= (x_1,x_2,x_3), \vy=(y_1,y_2,y_3)$ with $y_i = x_i + \D_i$ and $-4\le \D_i \le 4$.
This allows a maximum off-axis separation of $R=|\vx-\vy|=6.93$.  Four Laplacian eigenstates $\zeta_n(\vx)$, and
therefore eight operators $Q_\a(R)$ at each $R$, were used in the computation.

 Fig.\ \ref{combo} shows an example of our results
 for $\T_{nn}(R,T)$ at $R=5.83$.  We see that $\T_{11}(R,T)=1$ , independent of $T$, to a very high degree of
 precision.  This means that the ground state energy is $E_1(R)=0$, and it turns out that this result is obtained at
 all $R$, not just at $R=5.83$.  The $\T_{22}$ and $\T_{33}$ data are fit to an exponential
 \beq
             c_n(e^{-E_n T} + e^{-E_n(36-T)})  \ .
 \eeq
 It can be seen that the data for $\T_{22},\T_{33}$ fall off linearly on the logarithmic plot, with fits taken in
 the range $[2:7]$, and the slopes are slightly different, corresponding to a small
 energy difference between $E_2$ and $E_3$. Beyond $T=8$ or so there are rather large error bars, so those
 points are not included in the fit.

 \begin{figure}[htb]
 \includegraphics[scale=0.6]{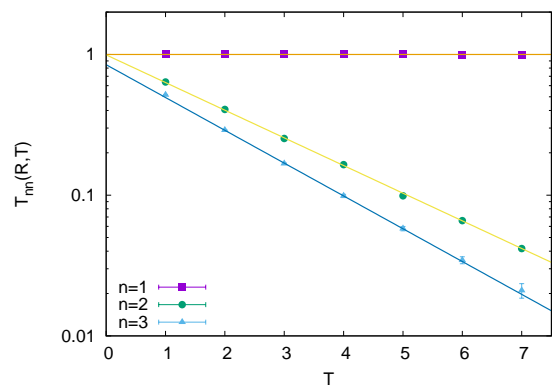}
 \caption{$\T_{nn}(R,T)$ vs.\ $T$ at $R=5.83$, for $n=1,2,3$ together with their best fits.  The corresponding
 energies $E_n(R)$ are extracted from the slope of exponential fits on a log plot.  This data was obtained
 on a $12^3 \times 36$ lattice volume at $\b=10.9$ and $\g=0.60$.}
 \label{combo}
 \end{figure}
 
    The energies $E_2(R), E_3(R)$ vs.\ $R$ are displayed in Fig.\ \ref{ERv12}.  Also shown is the mass of
 a static photon (blue line) at ${\g=0.6}$, as obtained in the previous section, along with the energy of
 the massive photon $\sqrt{m_{ph}^2 + (2\pi/12)^2}$ at the minimal non-zero momentum on a periodic lattice
 of length 12.  The ground state energy, not shown here, is at $E_1(R)=0$.
  
     The energy $E_2(R)$ of the first excited state above the ground state pretty nearly coincides with the
 photon mass.  So this state is easy to interpret: it is simply the ground state of the static charges  
 plus a static photon.  The second
 excited state, of energy $E_3(R)$, cannot be interpreted in that way, because (with the possible exception of one
 outlier) it lies above the one photon mass, but below the energy of a photon with minimal momentum.  So there
 are two possibilities.  One is that this state represents the energy of a static photon plus a small excitation, of
 energy $\D E = E_3 - m_{ph}$, of the field surrounding the static charge pair.  But in that case we might expect to see this energy  $\D E$ as the energy of the first excited state, i.e.\ an excitation without the massive photon, which is not seen.
 The second possibility is that $E_3$ represents an excitation energy of the static charges with no extra photons,
 which happens to lie above the threshold for photon production, and this seems more likely.  It of course implies
 that in the infinite volume limit, where the photon momentum can take on any value, this excitation is metastable, and could decay via  massive photon emission.  On the other hand, if the charges are very far apart, this option might be
 strongly suppressed, since the excitation energy of the field surrounding either charge is only half the total
 excitation energy, $\oh E_3$.
 
 \begin{figure}[htb]
 \includegraphics[scale=0.6]{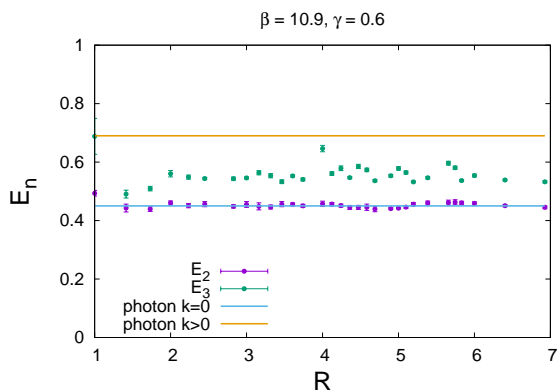}
 \caption{Energies $E_n(R)$ vs.\ $R$ for $n=2,3$ again at $\g=0.60$ on a $12^3 \times 16$ lattice volume.  The
 lower solid line represents the mass of a massive photon, and the upper line is the energy of a massive photon
 at the lowest possible momentum on this finite periodic lattice.}
 \label{ERv12}
 \end{figure}
 
     From the data, it appears that $\Psi_1(\vx,\vy)$ and $\Psi_2(\vx,\vy)$ are very close to the true ground state, and the ground state plus a massive photon respectively. We note in passing that this result required use of both types
of operators shown in \rf{Qn}.   Use of only the $Q_{2n-1}$ operators results in data that is not easily fit
by a single exponential.

 \subsection{Other Higgs couplings, volumes, and excitations}         
           
      There is no visible dependence on $R$ in $E_1(R)$, and very little in $E_2(R)$ in Fig.\ \ref{ERv12}.  However, we see significant scatter in the values shown for $E_3(R)$, which cannot be attributed to error bars in the exponential fit.    We conjecture that the reason for this scatter is that $|\Psi_3(\vx,\vy)\rangle$, while close to the true energy eigenstate, may contain enough (probably random) admixtures of higher excited states to show a rather small, mainly random dependence on $R$.   
                
           Figure \ref{ER16} is a comparison of the previous data in Fig.\ \ref{ERv12}, at volume $12^3 \times 36$, with data  obtained at the same $\b,\g$ parameters on a $16^3 \times 36$ lattice volume.  The data for $E_2$ coincide, while the data points for $E_3$ at the larger $L=16$ spatial extension are roughly compatible, sometimes coinciding, with the corresponding data points at $L=12$,  but there is much less scatter in the data at the larger volume.  This presumbably indicates a smaller admixture of higher excitations at large spatial volume.   A similar reduction in scatter, upon enlarging the spatial volume, is seen in the data for the energies $E_4(R)$ of the next excited state, displayed in Fig.\ \ref{ER4}. 

     At spatial extension $L=16$ the energy of a photon with smallest non-vanishing momentum is at $E = 0.597$, which is very close to to the $E_3$ data points.  If we had only the data on the $16^3 \times 36$ volume, we would probably conclude that $E_3$ corresponds to the ground state plus a massive photon of that kind.   But this interpretation is inconsistent with the data found on the smaller $12^3 \times 36$, and since increasing the volume has only a minimal effect on the values of $E_3$, we conclude that we are seeing the same state in both volumes, i.e.\ an excitation above
the ground state of the static ions, rather than a photon of non-zero momentum.   The data for $E_4$ is likewise almost
independent of the lattice spatial extension $L$, which argues against an interpretation involving one or more photons
of finite momentum.
           
 \begin{figure}[htb]
 \includegraphics[scale=0.6]{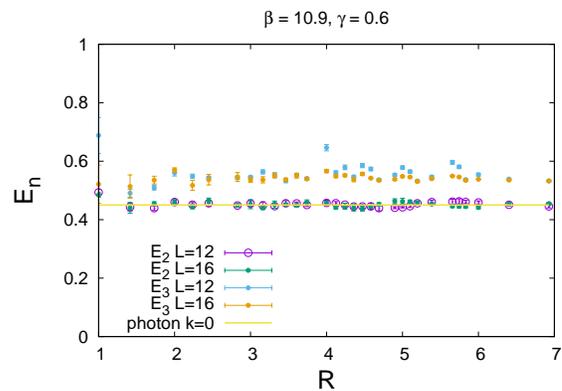}
 \caption{Comparison of $E_{2,3}(R)$ vs $R$ obtained on $12^3 \times 36$ and $16^3 \times 36$ lattice volumes at
 $\b=10.9, \g=0.60$.  The solid line represents the mass of a massive photon.}
 \label{ER16}
 \end{figure}      
 
 \begin{figure}[htb]
 \includegraphics[scale=0.6]{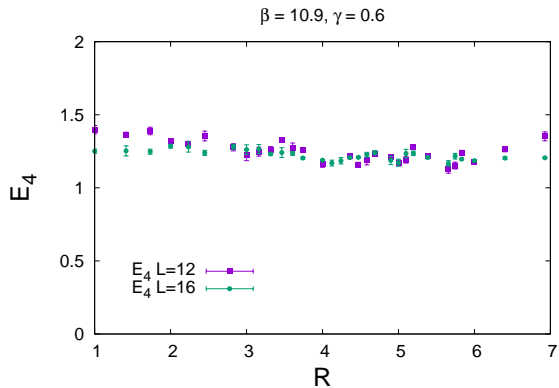}
 \caption{Same comparison of different volume data, as in Fig.\ \ref{ER16}, but this time for the excitation
 energy $E_4(R)$.}
 \label{ER4}
 \end{figure}   
 
       Figure \ref{ERg025} shows the results of the same calculation of $E_{2-4}$ on a  $16^3 \times 36$ volume, with the
same exponential fit in the range $T\in[2,7]$, but this time with $\g=0.25$.  The mass of the massive photon is
of course reduced (in lattice units) as $\g$ is reduced, as one expects, but the results are similar.  The next question is: how similar?

%and $\g=0.4$.  The mass of the massivephoton is of course reduced as $\g$ is reduced, as one expects, and we find also that the $E_3$ excitation energyis a little closer to the photon mass at $\g=0.4$.  Thus the precise ratio of $E_3/m_{ph}$ has some dependence on $\g$ or, equivalently, on the short distance cutoff (lattice spacing $a$) given a London penetration depth in physical units.

 \begin{figure}[htb]
 \includegraphics[scale=0.6]{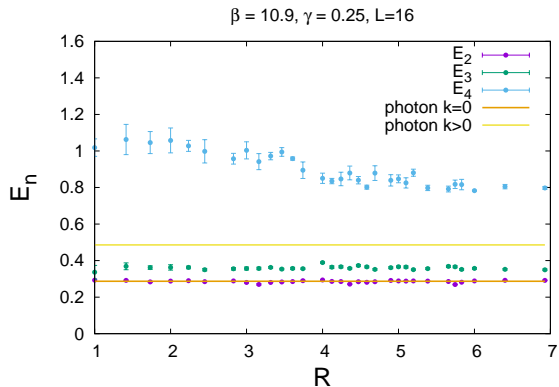}
 \caption{Energies $E_{2,3,4}(R)$, but this time at $\g=0.25$, with $\b=10.9$ and lattice volume $16^3 \times 36$. The lower and upper solid lines represent the energies of a static massive photon, and a photon of minimal momentum for this periodic volume.}
 \label{ERg025}
 \end{figure}

%\begin{figure}[htpb]
%\subfigure[~]{
 %\includegraphics[scale=0.5]{ER23g05.eps}
%}
%\subfigure[~]{
 %\includegraphics[scale=0.5]{ER23g04.eps}
%}
%\caption{}
%\label{ERg}
%\end{figure}

\subsection{Scaling}

    We have computed the photon mass and excitations $E_n$ in lattice units for two different $\g$ values, namely
$\g=0.25, 0.60$.  For a given London penetration depth $\lambda_L$ we can compute the lattice spacing, and
from there convert the charge separation $R$, photon mass, and excitation energies in physical units.  Let
$m_{ph}(\g)$ be the photon mass in lattice units found at a particular $\g$ value, and $M_{ph}(\g)=m_{ph}(\g)/a$ be the corresponding photon mass in physical units, where $a$ is the lattice spacing given in eq.\ \rf{spacing} . We then
have
\beq
         {M_{ph}(\g_1) \over M_{ph}(\g_2)} = \sqrt{\g_2 \over \g_1} 
                    {m_{ph}(\g_1) \over m_{ph}(\g_2)} 
\eeq
Inserting the computed values of $m_{ph}(\g=0.25) = 0.288(1)$ and $m_{ph}(\g=0.60)=0.446(3)$ in lattice units, we find the ratio of photon masses in physical units to be
\beq
 {M_{ph}(0.25) \over M_{ph}(0.60)} = 1.000(8)
\eeq

    If we choose a typical London penetration depth, say ${\l_L=50}$ nm, we can compare data for $E_n$ vs.\ $R$  in physical units, obtained at the two $\g$ values.  The result, for data taken on a $16^3\times 36$ lattice volume, is
shown in Fig.\ \ref{scaling}.  $E_1$ is consistent with zero for both $\g$ values, and the excitations $E_2, E_3$ essentially coincide.  Values for $E_4$ differ at small $R$, but appear to converge at larger $R$.  The overall conclusion is that the results in physical units do not really depend on $\g$, at least at larger charge separations. 

\begin{figure}[htb]
 \includegraphics[scale=0.6]{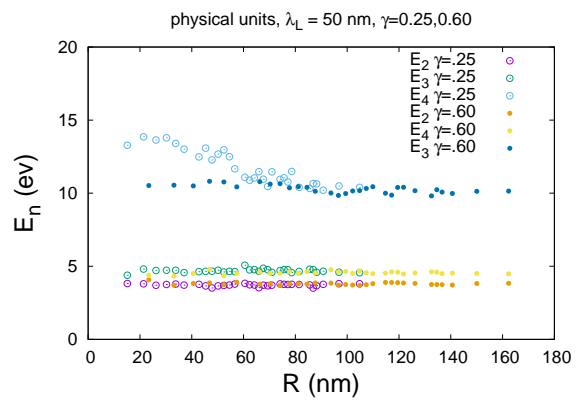}
 \caption{A test of scaling.  Using a fixed London penetration depth of $\l_L=50$ nm to set the scale, we convert both
 $R$ and $E_{2,3,4}(R)$ to physical units at both $\g=0.25$ and $\g=0.60$, with $\b=10.9$ fixed.}
 \label{scaling}
 \end{figure}  

\section{Discussion}

   The Ginzburg-Landau model is of course a simplified effective theory of superconductivity, presumably relevant at scales beyond the diameter of a Cooper pair.   But if the Ginzburg-Landau model is not misleading us, then the results presented here do suggest a possible experimental test via X-ray photoemission spectroscopy (XPS).  
   
    We have argued that the self-interacting field surrounding a static charge may have a spectrum of localized excitations.  In a certain sense, this has already been seen in normal metals.  The electric field of a static charge which is suddenly inserted into a normal metal is of course screened by electrons in the conduction band, and one can ask
whether these screening electrons themselves have a spectrum of excitations.  The answer appears to be yes, and this is the origin of the asymmetric lineshape in the core electron XPS spectrum, as pointed out long ago by
Doniach and Sunjic \cite{Doniach_1970} (for a simplified discussion, cf.\ \cite{huffner}).  When an electron is ejected from the valence band, this corresponds to a sudden appearance of an isolated charge in the metal, and the process of screening can leave the charge plus Fermi sea in one of a near-continuum of excitations slightly above the screened ground state, since very little energy is required to create electron-hole excitations near the Fermi level.  This results in the Doniach-Sunjic lineshape, which closely fits the XPS data.

   According to our results in the Ginzburg-Landau model, things may be a little different in the XPS spectrum of 
core electron emission in the superconducting phase of a conventional superconductor.  We have found a discrete
spectrum of excitations of the condensate, with energies on the order of mass of the massive photon.  This mass,
which in natural units is the inverse of the London penetration depth, depends on the metal, but in general is on the order of a few electron volts.  So in the superconducting phase we would expect to see a number of additional peaks in the core electron emission spectrum, separated from the main peaks by a few ev.  Unfortunately the experimental situation is a little different from our idealized setup of two static charges.  One of the charges, namely the hole left by the emitted core electron, is indeed static, and we would expect the condensate distribution around the hole to be no different from the idealized situation.  The emitted electron, however, is by no means static, which complicates making a precise
prediction for the location of the additional peaks.  If this complication is ignored, then 
given the Landau penetration depth $\l_L$, the positions of those peaks are calculable.  For, e.g., $\l_L=50$ nm,
the excitation energies for widely separated charges are the $E_3, E_4$ energies seen in Fig.\ \ref{scaling} at the
larger separations.   The relative heights and widths of these extra peaks depends on microscopic dynamics determining transition amplitudes and lifetimes, which are beyond the scope of the simplfied effective model.

   It appears that a comparison of XPS core electron spectra above and below the superconducting transition, for conventional (or, for that matter, high T$_c$) superconductors, has not yet been done.  At least, we have not been able to find a comparison of this type in the literature.  An experimental investigation along these lines could, for present purposes, be very helpful.

    An obvious question is whether quarks and leptons in the electroweak sector of the Standard Model, which is also a gauge Higgs theory, would have a spectrum of excitations analogous to what we find in the Ginzburg-Landau model.
If so, these would appear as ``elementary'' particles in their own right.  Unfortunately, in the absence of a lattice formulation of chiral gauge theory having a positive transfer matrix and a sensible continuum limit, we are unable to
make any predictions.  Should such a formulation ever become available, it would be possible to compute an excitation spectrum along the lines presented here.
   
\acknowledgments{We thank Peter Johnson and John Tranquada for helpful correspondence.   This research is supported by the U.S.\ Department of Energy under Grant No.\ DE-SC0013682.}     
   
\appendix*

\bigskip

\section{Numerical details}

     In all cases we have constructed, on each time slice, the Laplacian eigenstates $\zeta_n$ with the four lowest eigenvalues, leading to eight $|\Psi_n\rangle$ states, according to the procedures described above.  The Laplacian eigenstates were computed numerically via the Arnoldi algorithm, as implemented in the ARPACK software 
package.\footnote{https://www.caam.rice.edu/software/ARPACK/.} Data for 
 $[\T^T]_{\alpha \beta}(R)$, along with error bars for those values, were obtained by averaging the results of ten independent Monte Carlo simulations at $\g=0.25, 0.60$ on both $12^3 \times 36$ and $16^3 \times 36$ lattice volumes. In each simulation, 120 lattice configurations were generated after 5000 thermalizing sweeps, with data taking sweeps separated by 100 Monte Carlo update sweeps.  The numerical solution of the generalized eigenvalue problem was carried out by the Matlab {\tt eig} routine, which derived values and error bars for $\T_{nn}(R,T)$.  The determination of $\{E_n\}$ from fits to a single exponential, and corresponding error bars,
 were obtained from the gnuplot software.

     Fitting data points over a finite range raises the question of what is the ``best'' range, and how the answers would
differ if the range is slightly modified.  Although the lattice was large enough in the time direction to compute 
$\T_{nn}(R,T)$ up to $T=18$, in practice we found the statistical errors were significant for data points beyond $T=8$. Excluding points beyond those limits, and dropping the first data point (which would be the most susceptible to mixing from higher excitations) we can compare the values for $E_{2,3,4}$ using slightly different fitting intervals.
Typical results are shown in Table \ref{tab}, for $R=5.83$, and lattice size $16^3 \times 36$,  where $E_n^{0.25}, E_n^{0.60}$ refer to excitation energies computed at $\g=0.25$ and $\g=0.60$ respectively.  Error bars for $E_2, E_3$
were approximately one percent, and 3-4\% for $E_4$, which, given the small dependence on $R$, could indicate a slight overestimate of error bars on the fitted data points. The excitation energies shown in our figures were derived for a fitting interval of $R$ in the range $2-7$, but it is clear that choosing a different interval would not affect our conclusions.
 \begin{figure*}[htb]
\subfigure[~]  % caption for subfigure a
{ 
 \includegraphics[scale=0.6]{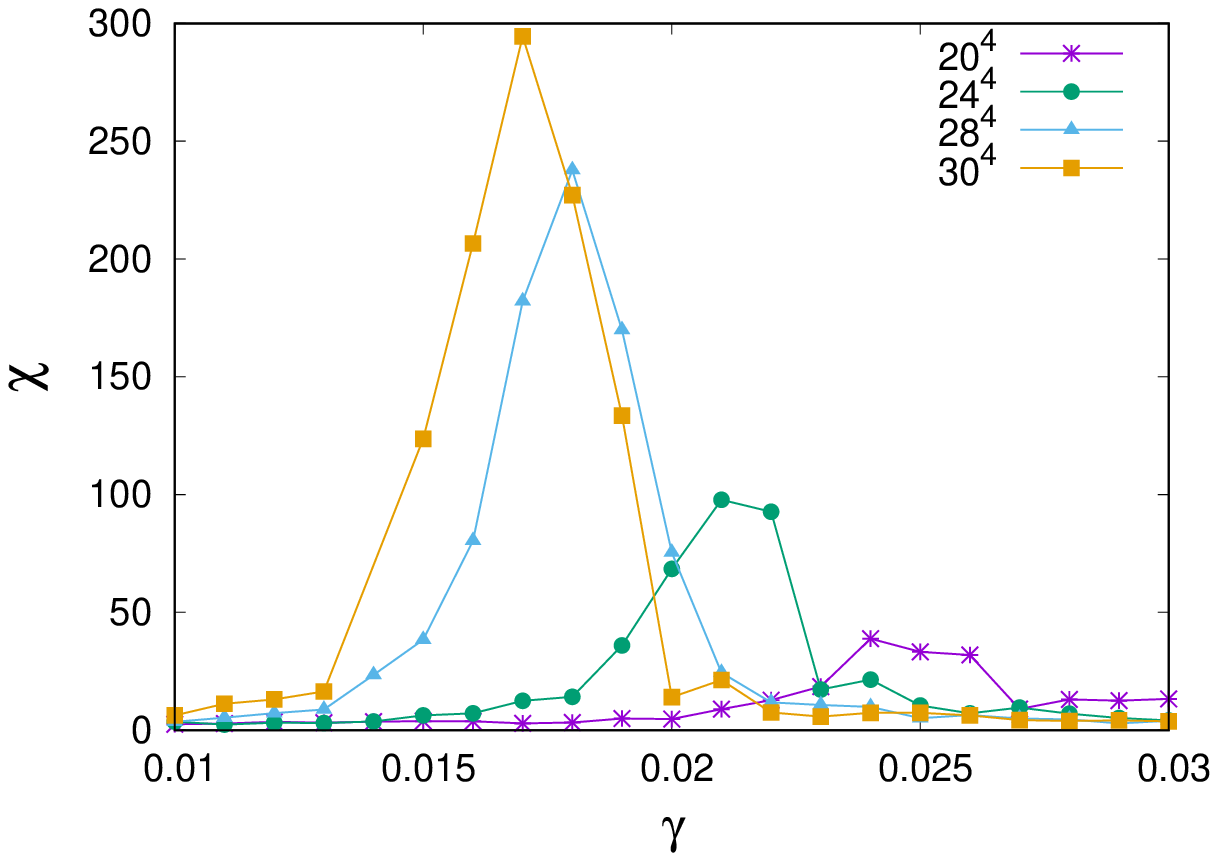}
 \label{sus}
}
\subfigure[~] 
{
 \includegraphics[scale=0.6]{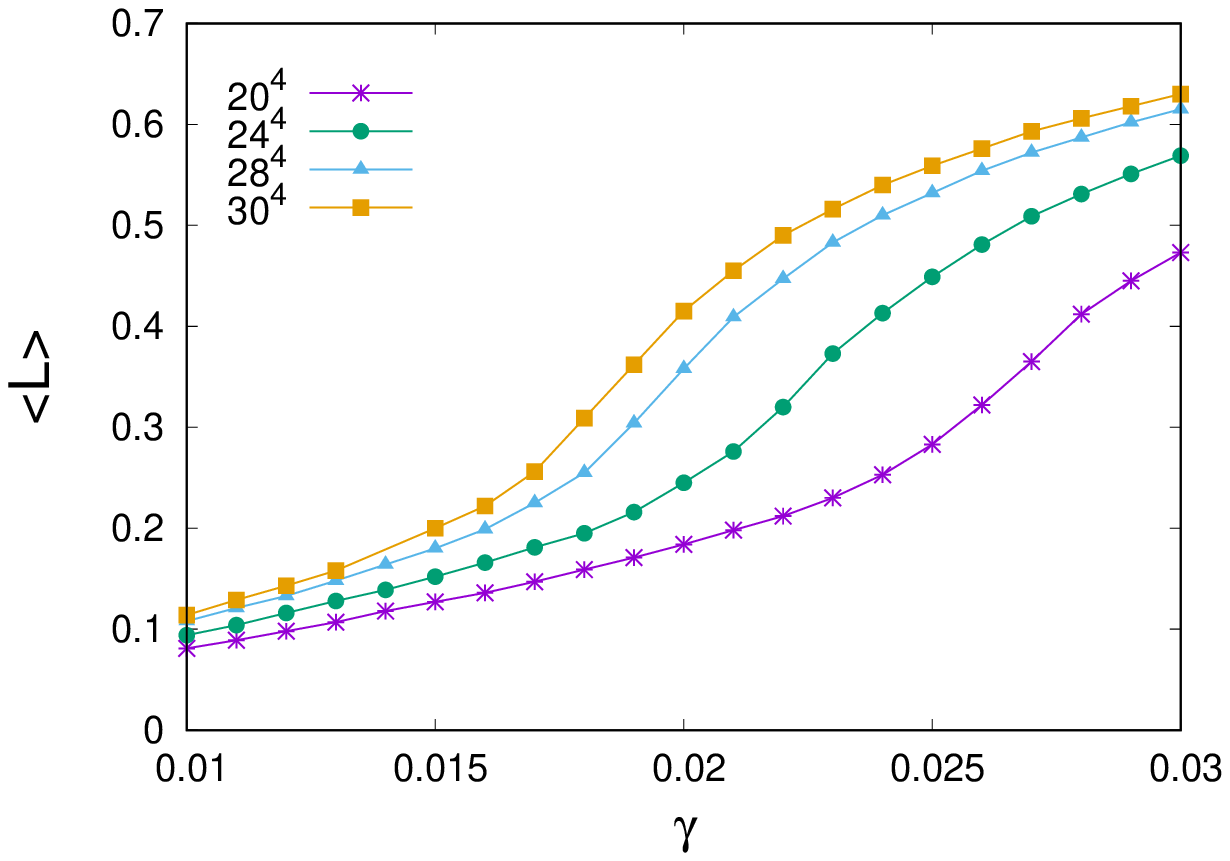}
 \label{link}
 }
 \caption{(a) Link susceptibility $\chi$ vs.\ $\g$; and (b) gauge-invariant link $\langle L \rangle$ vs.\ $\g$, on a variety of lattice volumes.}
 \label{phase}
 \end{figure*}
 
\begin{table*}[htb]
\begin{center}
\begin{tabular}{|c|c|c|c|c|c|c|} \hline
         fitting interval &  $E_2^{0.6}$ & $E_3^{0.6}$  & $E_4^{0.6}$ & $E_2^{0.25}$ & $E_3^{0.25}$  & $E_4^{0.25}$  \\  \hline
          2-4              &   0.472            &  0.543            &  1.18             &     0.288         &    0.363            &    0.83   \\
          2-5              &   0.472            &  0.540            &  1.13             &     0.286         &    0.357            &    0.81   \\
          2-6              &   0.453            &  0.538            &  1.13             &     0.282         &    0.354            &    0.81   \\
          2-7              &   0.453            &  0.537            &  1.13             &     0.282         &    0.353            &    0.82   \\
          2-8              &   0.453            &  0.537            &  1.13             &     0.282         &    0.350            &    0.81    \\                                   
\hline
\end{tabular}
\caption{Comparison of $E_{2,3,4}(R)$ at $\b=10.9$ and $\g=0.25, 0.60$ at $R=5.83$, obtained from fitting
the corresponding $\T_{nn}(R,T)$ vs.\ $T$ data in the different fitting intervals shown.}
\label{tab}
\end{center}
\end{table*}
 
      As mentioned in the text, at $\b=10.9$ we located a phase transition, presumably to the massless phase,
 at the rather low $\g$ value of $\g \approx 0.017$.  This was determined from inspection of the link susceptibility
 \bea
       \chi =  V (\langle L^2 \rangle - \langle L\rangle^2)
 \eea
 where $V$ is the lattice volume and
 \beq
       L = {1\over 3V} \sum_x \sum_{k=1}^3  \text{Re}[\phi^*(x) U^2_k(x) \phi(x+\hat{k})]
 \eeq
 with the result shown in Fig.\ \ref{sus}.  At each volume, data was taken on 1400 lattices separated by 100 sweeps, after 20,000 thermalizing sweeps. In previous work on the relativistic abelian Higgs model 
we have seen that the transition to the massless phase corresponds to a discontinuity (a ``kink'')  in the slope of $L$ in the infinite volume limit (cf.\ Fig.\ 1 in \cite{Matsuyama:2019lei}).  Although this is not as obvious in our Ginzburg-Landau data, there is some evidence of such a kink developing,  as the volume increases, in Fig.\ \ref{link}.

\bibliography{sym3}

\end{document}